\documentclass[preprint]{revtex4-1}
\usepackage[pdftex]{graphicx}
\usepackage{filecontents}
\usepackage{amsmath}
\pdfoutput=1
\usepackage{graphicx}
\usepackage{subfig}
\usepackage{color}

\begin{document}

	\begin{abstract}
		A common use for atomic force microscopy is to quantify local forces through tip-sample interactions between the probe tip and a sample surface. The accuracy of these measurements depends on the accuracy to which the cantilever spring constant is known. Recent work has demonstrated that the measured spring constant of a cantilever can vary up to a factor of two, even for the exact same cantilever measured by different users on different microscopes. Here we demonstrate that a standard method for calibrating the spring constant (using the oscillations due to thermal energy) is susceptible to ambient noise, which can alter the result significantly. We demonstrate a new step-by-step method to measure the spring constant by actively driving the cantilever to measure the resonance frequency and quality factor, giving results that are unaffected by acoustic noise. Our method can be performed rapidly on any atomic force microscope without any expensive additional hardware.
	\end{abstract}
	\title{Eliminating the effect of acoustic noise on cantilever spring constant calibration}
	\author{Aaron Mascaro}
	\email{Author to whom correspondence should be addressed: mascaroa@physics.mcgill.ca}
	\author{Yoichi Miyahara}
	\author{Omur E. Dagdeviren}
	\author{Peter Gr{\"u}tter}
	\affiliation{Department of Physics, McGill University, 3600 rue University, Montreal,  Qu\'{e}bec H3A2T8, Canada}
	
	\maketitle
	
	The atomic force microscope (AFM) has become an invaluable tool across many areas of materials science research due to its ability to probe structural and electrical properties of materials with extremely high spatial resolution. Modern AFMs rely on a micro-fabricated sharp probe tip protruding from the end of a cantilever beam to sense exceptionally small forces \cite{albrecht1990microfabrication,ohnesorge1993true,butt2005force,raiteri2001micromechanical,giessibl1995atomic}. In many experiments the interaction force itself is to be measured, which is generally done by measuring the change in the mechanical status of the cantilever (static deflection, oscillation amplitude, or change in resonance frequency) as it interacts with the surface \cite{li2007ultra,raiteri2001micromechanical,noy1997chemical,wiggins2006high,balke2014exploring}.\\

	Independent of the operation mode of the AFM, the spring constant of the cantilever needs to be known to convert the measured cantilever response to units of force, which can then be used to quantify the tip-sample interaction strength \cite{giessibl2001direct,sader2004accurate,butt2005force,garcia2002dynamic}. There are several methods currently used to quantify spring constants including the method of Cleveland et al. \cite{cleveland1993nondestructive} where the cantilever's resonance frequency ($\omega_0$) is measured before and after adding known masses to the end of the cantilever, and Sader's method \cite{sader1999calibration}, which requires knowledge of the cantilever's resonance frequency ($\omega_0$), quality factor (\emph{Q}), plan-view dimensions (length \emph{L} and width \emph{b}), and the viscous medium the cantilever resides in (typically air). Due to its non-invasive nature, Sader's method has been widely adopted across commercial AFM systems for cantilever spring constant calibration. A common implementation of Sader's method is to measure the power spectral density (PSD) of the cantilever's deflection to observe the thermal oscillations, which can then be used to extract both the quality factor and resonance frequency, although Sader's method is fundamentally agnostic as to how the quality factor and resonance frequency are actually measured. Sader et al. have recently shown that the variation on these parameters obtained by fitting the measured thermal PSD can lead to differences of up to a factor of 2 in the spring constant obtained using Sader's method by different users on different microscopes even for the exact same cantilever \cite{sader2016virtual,te2011interlaboratory}. This technique assumes that thermal fluctuations are the sole driving force acting on the cantilever, which results in spectrally white multiplicative noise \cite{landau2013course}. This may be true in many cases, however, we demonstrate that additional noise sources such as ambient acoustic noise can cause the overall driving force to deviate from white Gaussian noise, which can alter the values obtained by fitting the measured PSD to that of a damped driven harmonic oscillator driven by Brownian noise. Furthermore, we demonstrate that by actively driving the cantilever we can obtain reliable measurements of the resonance frequency and quality factor that are impervious to increased ambient acoustic noise levels. \\

\begin{figure}[ht]
		\includegraphics[width=0.6\linewidth]{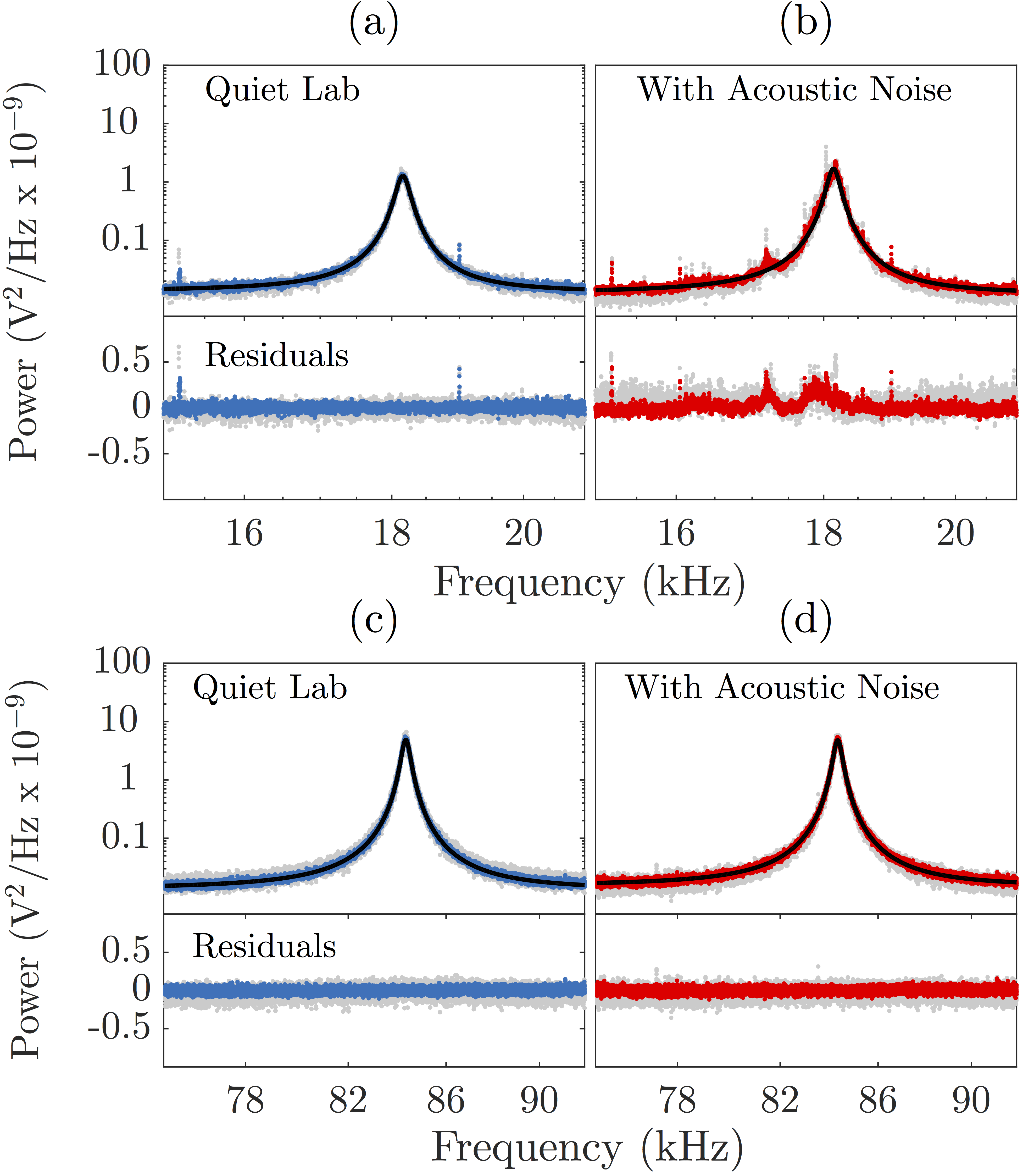}
		\caption{Measured PSD around the cantilever resonance for Cantilever Type~1~-~A: \textbf{(a)} without acoustic noise and \textbf{(b)} with ambient acoustic noise; and Cantilever Type~2~-~A: \textbf{(c)} without acoustic noise and \textbf{(d)} with ambient acoustic noise. Coloured data are the average of 5 independent measurements of 50 spectra averaged together, while the grey data shows one such measurement of 50 spectra averaged. Black lines are the fits to the logarithm of Equation \ref{eq:transfer_func_fit} \cite{Note1}.}
		\label{fig:spectra}
	\end{figure}

	Figures \ref{fig:spectra}(a) and \ref{fig:spectra}(c) show typical frequency spectra of the thermal oscillation peaks of two different cantilevers (Type~1: OPUS 4XC-NN-A, and Type~2: OPUS 4XC-NN-B) obtained by recording the AFM deflection signal at a sample rate of 2.5MHz for 2.5s, taking a fast-fourier transform (FFT), and then averaging 50 times (as per the recommended procedure of Sader et. al \cite{sader2012effect}). Modelling the cantilevers as damped driven harmonic oscillators, the frequency spectra of the oscillation peaks are given by:
	
	\begin{equation}
		F(\omega,\bar{\alpha}) = \frac{\alpha_1/\omega_0^2}{(1 - (\omega/\omega_0)^2)^2 + (\omega/\omega_0 Q)^2} + \alpha_2
		\label{eq:transfer_func_fit}
	\end{equation}
	\\
	where $F(\omega,\bar{\alpha})$ is the power spectral density (PSD) (in V$^2$/Hz or m$^2$/Hz), $\alpha_1$ is the amplitude, and $\alpha_2$ is the baseline noise level.  A least-squares fit to the logarithm of Equation \ref{eq:transfer_func_fit} is shown as the solid black line in each panel of Figure \ref{fig:spectra}  \footnote{Since the thermal noise is multiplicative, taking the logarithm of both sides of Equation \ref{eq:transfer_func_fit} removes the weighting of the squared errors in the least-squares minimization procedure and results in residuals that are zero-centered.}, where the window sizes are large compared to the spectral width of the lorentzians (corresponding to a normalized window size of $\beta \approx$17 as defined by Sader et al. \cite{sader2014uncertainty}), which results in small uncertainties on the fit parameters.\\
	
	To study the effect of ambient noise on the measurements, a speaker (Motorola J03 type) was connected to the output of a function generator (Agilent 33220a) outputting white noise with a bandwidth of 9MHz and placed near the AFM. The frequency spectra for Cantilevers Type~1~-~A and Type~2~-~A with ambient acoustic noise are shown in Figures \ref{fig:spectra}(b) and \ref{fig:spectra}(d). To preclude effects of slowly changing extrinsic variables that could affect the measurements, the quiet and noisy measurements were done in an alternating fashion.\\
	
\begin{center}
		\begin{figure}[ht]
			\includegraphics[width=0.9\linewidth]{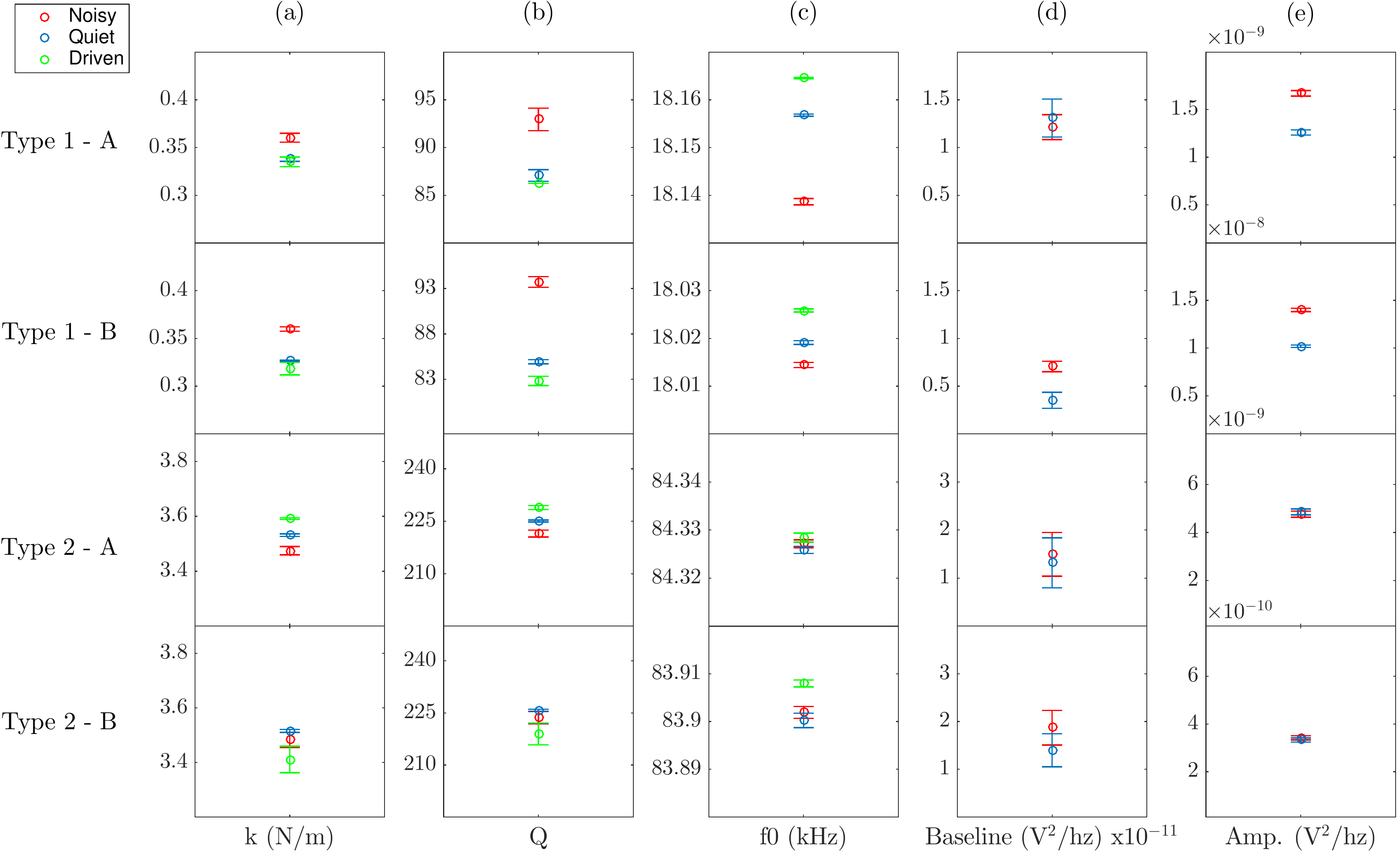}
			\caption{\textbf{(a)} Spring constants obtained from Equation \ref{eq:sader_eq}. \textbf{(b)} quality factors, \textbf{(c)} resonance frequencies ($\omega_0/2\pi$), \textbf{(d)} baseline noise levels ($\alpha_2$) and \textbf{(e)} peak amplitudes ($\alpha_1$) obtained from fitting the `thermal' oscillation PSD measurements of 4 different cantilevers with (red) and without (blue) ambient acoustic noise to Equation \ref{eq:transfer_func_fit}. Results for the driven-calibration method described in the text are shown in green. Error bars are the uncertainty on the mean from 5 measurements on each cantilever.}
			\label{fig:all_data}
		\end{figure}
	\end{center}

	Equation \ref{eq:transfer_func_fit} was used to determine the quality factor, resonance frequency, baseline noise level, and  amplitude of four different cantilevers, two of Type~1 with resonance frequencies in the audio range ($\sim$20kHz), and two of Type~2 with resonance frequencies well into the ultrasonic range ($\sim$80kHz). These results are shown in Figures \ref{fig:all_data}(b)-(c) where the error bars are the standard deviation of the mean for 5 independent measurements of each cantilever. The `noisy' data (red data points in \ref{fig:all_data}) are values obtained from fitting the frequency spectra with ambient acoustic noise as described.\\
	
	The spring constant for a rectangular cantilever can be directly calculated by:
	
	\begin{equation}
	k_n = 0.1906\rho b^2 L Q \: \Gamma_i(\omega_0)\omega_0^2
	\label{eq:sader_eq}
	\end{equation}
	
	where the prefactor (0.1906) comes from the normalized effective mass and $\Gamma_i$ is the imaginary component of the hydrodynamic function \cite{sader1999calibration}. The spring constants for all four cantilevers were calculated using Equation \ref{eq:sader_eq} and are shown in Figure \ref{fig:all_data}(a).\\ 

\begin{figure}[ht]
		\includegraphics[width=0.32\linewidth]{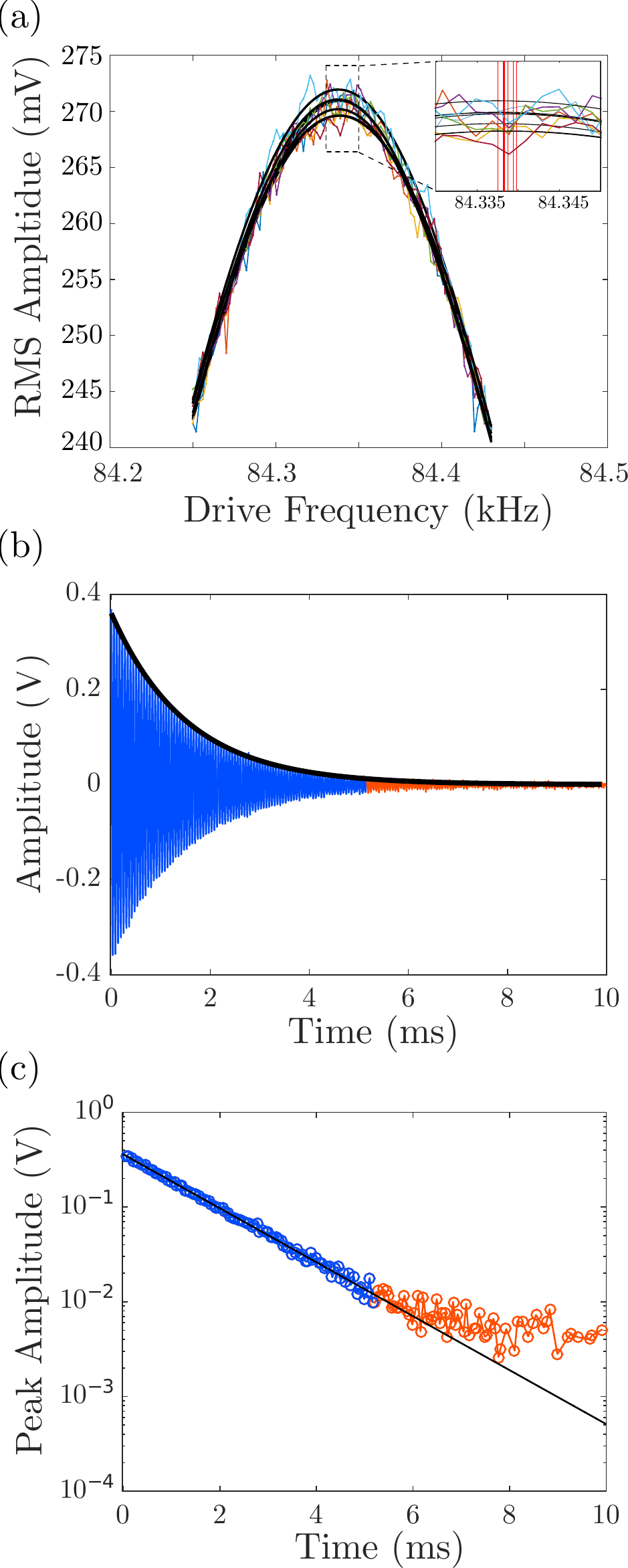}
		\caption{\textbf{(a)} Driven response amplitude (RMS) of Cantilever Type~1~-~A across the resonance frequency showing curves fitted to Equation \ref{eq:transfer_func_fit}. Inset shows a closeup where the red vertical lines mark the resonance frequency obtained from each fitted curve. \textbf{(b)} AFM deflection signal directly after turning off the driving force (at t=0) along with the fit to Equation \ref{eq:ringdown} for multiple measurements on cantilever Type~1~-~A. \textbf{(c)} Driven response amplitude (RMS) of cantilever Type~1~-~A across the resonance frequency showing curves fitted to \ref{eq:transfer_func_fit}. Inset shows a closeup where the red vertical lines markt the resonance frequency obtained from each fitted curve.}
		\label{fig:sweep_ringdown_log}
	\end{figure}

	Another method of measuring the resonance frequency is to drive the cantilever using a sine wave and sweeping its frequency. This can be done using a piezo-acoustic drive, which is susceptible to the non-flat transfer function of the system \cite{labuda2011decoupling}. Since the quality factor is equal to $f_0$/FWHM (where $f_0$ is the resonance frequency in Hz and FWHM is the full width at half-max of the resonance peak) and typical quality factors are $\sim$200 for the $\sim$80kHz cantilevers used in this experiment, the frequency span required to measure \emph{Q} from a driven spectrum would be at least 1kHz. Thus, determing \emph{Q} from a driven cantilever response by fitting the peak would be highly susceptible to transfer function irregularities and/or spurious resonances within this $\sim$1kHz range. We can, however, measure the resonance frequency of the cantilever very accurately by sweeping over a small frequency window and fitting the response to Equation \ref{eq:transfer_func_fit} as the transfer function should have a minimal impact as long as the frequency span is small enough. Multiple measurements on Cantilever Type~2~-~A are shown in Figure \ref{fig:sweep_ringdown_log}(a) along with their fitted curves (black lines). The inset shows the accuracy of the fits, each measurement is within 2Hz of the mean and the uncertainty on the mean is under 1Hz. The measurement is unaffected by adding acoustic noise (i.e. the results with and without noise are the same).\\
	
	To measure the quality factor using a driven technique we can simply record the ringdown time since the quality factor of a damped harmonic oscillator is defined as the oscillator's stored energy divided by the energy lost per oscillation cycle (times 2$	\pi$). This was performed by driving the cantilevers at the resonance frequency previously measured and suddenly turning off the driving force. By directly recording the AFM deflection signal we can observe the oscillation amplitude decreasing, as shown in Figure \ref{fig:sweep_ringdown_log}(b). The peak values can be easily extracted using a simple peak-finding algorithm, and they decrease exponentially over time, given by:
	
	\begin{equation}
		y = Ae^{-t/\tau}
		\label{eq:ringdown}
	\end{equation}
	
	where $\tau$ is the decay time constant and \emph{A} is the exponential prefactor. The quality factor is related to the decay time constant by:
	
	\begin{equation}
		Q = \tau f_0 \pi
		\label{eq:qfactor}
	\end{equation}

	In fitting the peak amplitudes to Equation \ref{eq:ringdown}, one has to be aware of the effect of the non-zero noise floor of the measurement device. This becomes apparent when plotted on a log-log scale: as the peak values approach the noise floor they begin to deviate from the expected straight-line behaviour, as shown by the red data points in Figure \ref{fig:sweep_ringdown_log}(c). This can easily be corrected for by simply measuring the noise-floor, which we define as the peak-to-peak noise on the deflection signal with the drive turned off, and then only including peak values greater than this value in the fit. These are shown in blue in Figure \ref{fig:sweep_ringdown_log}(c), while the red data points were not included in the fit.\\ 

	The results for each cantilever from the sweep and ringdown measurements are shown in green in Figure \ref{fig:all_data}. The same values were obtained with and without ambient acoustic noise.\\
	
	Although we used the ringdown method for quality factor measurements, there exist other driven techniques to extract the quality factor as well, including by taking the derivative of the measured phase vs. frequency data. This quantity is related to the quality factor by $\frac{d\phi}{d\omega}|_{\omega = \omega_0} = 2Q/\omega_0$ where $\phi$ is the oscillator phase with respect to the drive signal \cite{marion2013classical}. The main drawback of this technique is the numerical derivative that must be computed, which is widely known to greatly amplify noise present in the data. This technique therefore requires significant averaging in order to obtain reliable results, and in addition it is also susceptible to transfer function irregularities as with any measurement where the drive frequency is swept. The ringdown technique, on the other hand, requires excitation at a single frequency and is thus impervious to effects related to the mechanical transfer function.\\	
	
	As can be observed in Figure \ref{fig:spectra}, ambient acoustic noise can affect the measured PSD. This is immediately apparent in the case of the audio-frequency range cantilever Type~1-~A (Figures \ref{fig:spectra}(a) and (b)), while the spectrum for the ultrasonic frequency-range cantilever Type~2~-~A is visually indistinguishable with and without ambient acoustic noise (Figures \ref{fig:spectra}(c) and (d)). As shown in Figure \ref{fig:all_data}(a), the spring constant obtained from fitting the thermal PSD may be systematically overestimated by 10\% in some cases (Type~1 Cantilevers), while in others it may be underestimated (Type~2~-~A), and in the best case there is no observed difference (Type~2~-~B). Using the driven techniques we described, however, yeilded spring constants consistent with those obtained from the quiet thermal spectrum measurements and were unaffected by acoustic noise.  \\
	
	The Type~2 cantilevers have both larger spring constants and resonance frequencies in the ultrasonic range. The acoustic noise generated by the speaker does extend well into the ultrasound, however atmospheric attenuation at higher frequencies is known to be severe \cite{lawrence1982measurements}. Thus, as expected, the stiffer, higher frequency cantilevers are less affected by ambient acoustic noise, but not impervious to it. To understand why the fit results differ for Cantilever Type~2~-~A even though there are no clear visual differences in the data, it is instructive to look at the variance of the residuals (\emph{R}) since the residuals are proportional to the logarithm of the noise. Taking $\textrm{Var}[10^R]$ where $R = \log_{10}[y] - \log_{10}[F(\omega,\bar{\alpha})]$ (i.e. the logarithm of the data minus the logarithm of the fit function, Equation \ref{eq:transfer_func_fit}, which gives a unitless quantity) we can compare how `noisy' the residuals are. For Cantilever Type~1~-~A the variances are: $3.0 \pm 0.3 \times 10^{-2}$ for the quiet data and $5.8 \pm 0.4 \times 10^{-2}$ for the noisy data, while for Cantilever Type~2~-~A the variances are: $2.06 \pm 0.04 \times 10^{-2}$ for the quiet data and $2.42 \pm 0.04 \times 10^{-2}$ for the noisy data. In both cases the residuals are significantly noisier when the acoustic noise is on. \\ 
	
	This discrepancy is fundamentally due to the fact that the observed spectrum is not always thermally limited; there can be contributions from various sources of detection noise (e.g. optical shot noise), electronic noise, and mechanical vibrations (e.g. acoustic noise from vacuum pumps). The former have been investigated comprehensively for optical beam deflection systems such as the one used here~ \cite{butt1995calculation,fukuma2009wideband,giessibl1999calculation,labuda2011noise,gittes1998thermal}, while the effect of mechanical vibrations on the thermal oscillations of tuning forks have been discussed in brief \cite{welker2011application}. Since the energy of the thermal oscillations is so small, even a small amount of mechanical noise (acoustic or otherwise) can have a non-negligible effect and lead to deviations from a spectrally white driving force. This is evident in the residuals plotted in Figure \ref{fig:spectra}(b). The deviation from a Lorentzian is due to the acoustic energy being converted into mechanical oscillations of the cantilever around the cantilever's resonance frequency. Note that the mechanical transfer function of an AFM system is not flat in frequency due to many unavoidable non-linear mechanical couplings existing between the different microscope components. It is these couplings that lead to frequency dependent phase shifts described and measured in Ref. \cite{labuda2011decoupling}. The exact mechanism by which acoustic noise presents in the cantilever deflection PSD is expected to be highly dependent on the geometry of the microscope and the noise source itself.  By actively driving the cantilevers, however, the energy of the mechanical oscillations can be increased well above the noise floor making them insensitive to ambient acoustic noise. \\

	As we have shown, ambient acoustic noise can introduce systematic errors into thermal measurements of cantilever quality factors, which can propagate to errors in calculate spring constants. This effect is especially pronounced for cantilevers with resonance frequencies in the audio range ($\textless$20kHz), but can also be present for cantilevers with resonance frequencies well above this. By actively driving the cantilever to measure the resonance frequency and quality factor, the effect of acoustic noise can be mitigated. The quality factor can reliably be measured by recording the ringdown directly and fitting this to a decaying exponential. The fit should be done such that only data above the noise floor is included. This procedure results in highly reproducible measurements that can be used to calculate the spring constant of a cantilever using standard techniques and precludes systematic errors due to ambient acoustic noise, which may contribute to observed differences in cantilever spring constants obtained on different atomic force microscopes and/or by different users.\\

	The authors acknowledge financial support from the Natural Sciences and Engineering Research Council of Canada, and the Fonds de recherche du Quebec - Nature et Technologies. 

	\bibliographystyle{apsrev4-1}
	\bibliography{bib}
	
	\pagebreak

\end{document}